\documentclass[fleqn,usenatbib]{mnras}

\pdfoutput=1

\usepackage{newtxtext,newtxmath}

\usepackage[T1]{fontenc}
\usepackage{ae,aecompl}

\usepackage{color}
\usepackage{ulem}

\usepackage{graphicx}	
\usepackage{epstopdf}
\usepackage{amsmath}	
\usepackage{amssymb}	
\usepackage{caption}

\newcommand\tff{t_{\rm ff}}

\newcommand\Msun  {{\rm M}_\odot}

\newcommand\kms{\rm \, km ~ s^{-1}}

\newcommand{\ppcc}{{\rm cm}^{-3}}

\newcommand\beq{\begin{equation}}
\newcommand\eeq{\end{equation}}

\newcommand{\mG}{\mu {\rm G}}

\def\alamenos#1{$^{-#1}$}

\def\apj{ApJ}
\def\apjl{ApJL}
\def\aap{A\& A}

\def\mnras{MNRAS}

\def\apjs{ApJS}

\def\mnras{{MNRAS}}
\def\pasj{{PASJ}}

\title[Are fibres real objects?] {Are fibres in molecular cloud filaments real objects?} \author[Zamora-Avil\'es, Ballesteros-Paredes \& Hartmann]
      {Manuel Zamora-Avil\'es,$^{1,2}$ \thanks{E-mail:
          manuelaz@umich.edu, m.zamora@crya.unam.mx}, Javier
        Ballesteros-Paredes$^{2,3}$, Lee W. Hartmann$^1$
\\
$^{1}$Department of Astronomy, University of Michigan,  500
           Church Street, Ann Arbor, MI 48105, USA \\
$^{2}$Centro de Radioastronom\'ia y Astrof\'isica,
            Universidad Nacional Aut\'onoma de M\'exico,
            Apdo. Postal 72-3 (Xangari), Morelia,\\
            Michoc\'an 58089, M\'exico \\
$^{3}$Zentrum f\"ur Astronomie der Universit\"at Heidelberg, Institut f\"ur Theoretische Astrophysik, Albert-Ueberle-Stra{\ss}e 2, D-69120 \\
Heidelberg, Germany
}

\date{Accepted XXX. Received YYY; in original form ZZZ}

\pubyear{2017}

\begin{document}
\label{firstpage}
\pagerange{\pageref{firstpage}--\pageref{lastpage}} \maketitle

\begin{abstract}

We analyse the morphology and kinematics of dense filamentary structures 
produced in a numerical simulation of a star--forming cloud of $1.4 \times 
10^4 \, \Msun$ evolving under their own self--gravity in a magnetized media. 
This study is motivated by recent observations of velocity--coherent 
substructures (``fibres") in star-forming filaments. We find such ``fibres" 
ubiquitously in our simulated filament. We found that a fibre in one 
projection is not necessarily a fibre in another projection, and thus, 
caution should be taken into account when considering them as real objects. 
We found that only the densest parts of the filament ($\sim$30\% of the 
densest volume, which contains $\sim$70\% of the mass) belongs to fibres 
in 2 projections. Moreover, it is quite common that they are formed by 
separated density enhancements superposed along the line of sight. 
Observations of fibres can yield insight into the level of turbulent 
substructure driven by gravity, but care should be taken in interpreting 
the results given the problem of line of sight superposition. We also 
studied the morphology and kinematics of the 3D skeleton (spine), 
finding that subfilaments accrete structured material mainly along 
the magnetic field lines, which are preferentially perpendicular to 
the skeleton. The magnetic field is at the same time dragged by the 
velocity field due to the gravitational collapse.

\end{abstract}



\section{Introduction} \label{sec:intro}

Although filamentary structure in star-forming molecular clouds and
young stellar populations has long been recognized
\citep[e.g.,][]{schneider1979,hartmann2002,nutter2008}, as well as the
possible importance of filamentary geometry on scales of gravitational
fragmentation \citep{larson1985,larson2005}, the subject has attracted
increasing attention following the detection of ubiquitous filamentary
structure in far-infrared imaging with the {\it Herschel Space 
Telescope}, generally associated with recent and/or ongoing star
formation \citep{andre2010,mensh2010,arzoumanian2011,hacar2011}.

Recently, kinematic investigations, most notably of the B211/213 cloud
in Taurus, have provided evidence for velocity-coherent substructures
within filaments \citep[called ``fibres";][]{hacar2013}.  This led
\cite{tafalla2015} to propose the ``fray and fragment" scenario for
filament and star formation, in which supersonic flows form a
filamentary density enhancement. In this scenario, the filament then
splits into intertwined velocity-coherent fibres as a result of
turbulent motions and gravitational driving (the ``fray" step).
Finally, some fibres acquire enough mass to gravitationally fragment
into cores and stars. Some theoretical support for this idea has been
provided by \citet{clarke2017}, who numerically simulated the
evolution of a cylinder in a converging, cylindrical flow with
additional imposed turbulence.  On the other hand, large-scale cloud
simulations by \cite{smith2016} indicated that subfilaments, which
they interpret as possible fibres, formed first, and then gather
together to form denser main filaments, suggesting a ``fray and
gather'' scenario.

There have been many theoretical investigations of filaments, including 
analytic studies of idealized filaments, both non-magnetized
\citep{larson1985,larson2005} and magnetized
\citep{hennebelle2003,tilley2003,seifried2015}, as well as accreting
filaments \citep[which must occur to form filaments in the first 
place;][]{fischera2012,heitsch2013}.  While these studies have
provided important general insights into filament properties, to study
the relation of fibres to filaments it is necessary to conduct
large-scale simulations of turbulent clouds evolving under self
gravity \citep[e.g.,][]{jappsen2005,gomez2014,smith2014,moeckel2015},
as these simulations produce filaments with substructure that
otherwise need to be put in by hand in smaller-scale calculations.

In this contribution we present simulations of an initially turbulent 
cloud evolving under gravity, similar to those of \citet{smith2016} 
but with the inclusion of magnetic fields, which are expected to 
enhance the formation of filaments perpendicular to the field 
\citep{hennebelle2013}. In order to compare more directly with 
previous observational work, we map our data into 
Position--Position--Velocity (PPV) data cubes, and apply a 
Friends--of--Friends in velocity algorithm (FoFv) similar to the 
Friends-in-Velocity (FIVe) algorithm used by \cite{hacar2013}
to identify fibres. However, in order to assess the physical 
coherence of the fibres and to compare with previous numerical work, 
we apply {\tt DisPerSE} \citep{disperse1, disperse2}, an algorithm 
that searches for the ``skeletons" of the density field. 

We find that velocity-coherent fibres as observed by 
\cite{hacar2013,tafalla2015} are ubiquitous in our PPV maps. 
However, fibres in one PPV projection share little correspondance
with fibres in another projection, implying that fibres are the 
result of superposition,
rather than been actual physical, spatially and velocity coherent 
structures \citep[e.g.,][]{BM02,gammie2003}. We also find that, 
although the filaments may be moving in different directions as a 
whole, in the local frame of the skeleton they are accreting mass 
from their environment.  

\begin{figure*}
    \includegraphics[width=1\hsize]{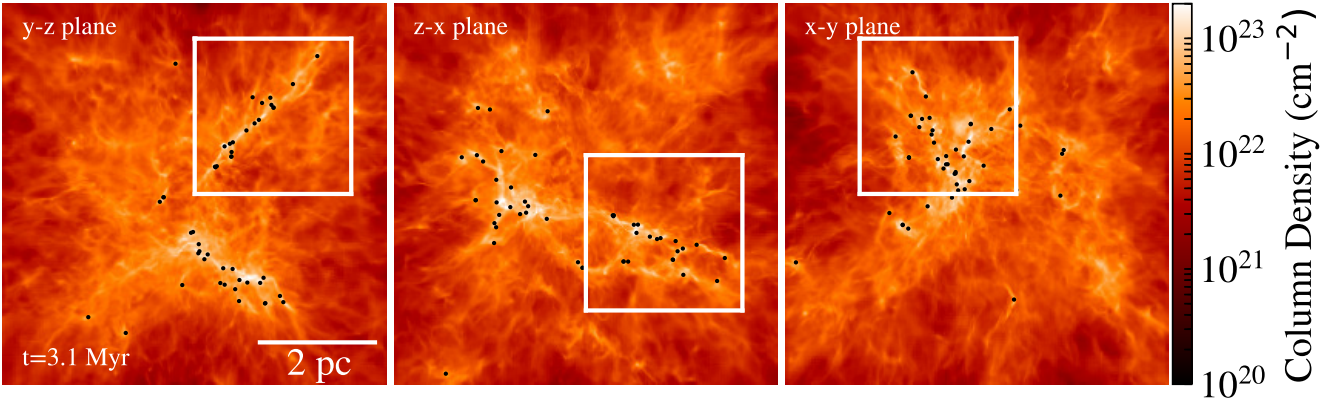}
	\caption{Column density maps integrated along the $x$, $y$, and 
	$z$--directions (left, middle, and right panels, respectively) of 
	the central (7.25 pc) cloud at $t=3.1$ Myr ($\sim 0.9 \, \tff$). 
	The black dots correspond to the positions of the sink particles. 
	The white squares in each panel delimit the sub--box (of $\sim$ 2.7 
	pc per side) enclosing the studied filament.}
	\label{fig:cloud}
\end{figure*}

Our results emphasize the importance of projection in spatially
superposing physically-distinct velocity components. Furthermore, 
our models suggest that fibres in general are not a result of filament 
splitting.  The inflows 
are the result of gravitational acceleration toward the growing filament. 
A ``hub-spoke" geometry, with subfilaments accreting toward mass
concentrations (cores, sinks) within filaments, often occurs. 

The plan of the paper is as follows. In \S \ref{sec:numerics} and 
\S \ref{subsec:ppv} we describe the numerical methods and we present 
our results in \S \ref{sec:results}. In \S \ref{sec:discussion} we 
compare our findings with those of \cite{smith2016} and 
\citet{clarke2017}, and discuss the implications for the ``fray and 
fragment" scenario. Finally, we summarize the results in \S 
\ref{sec:conclusion}.

\section{Numerical Simulations} \label{sec:numerics}

We use the Eulerian adaptive mesh refinement FLASH (v2.5) code
\citep{Fryxell+00} to perform three-dimensional, self gravitating,
magnetohydrodynamic (MHD) simulations of a hierarchically collapsing
molecular cloud in an isothermal regime. The ideal MHD equations are 
solved using the MHD HLL3R solver \citep[see e.g.,][]{Waagan+11}, which 
present a good compromise between accuracy and robustness for highly 
supersonic astrophysical problems, such as those studied here. 
We impose periodic boundary conditions for the MHD and 
isolated ones for gravity.

In order to follow the development of high density regions, we dynamically 
refine according to the 
Jeans criterion, which aims to prevent spurious fragmentation, 
by requiring that the local Jeans length be resolved 
by at least four grid cells \citep{Truelove+97}. Once the maximum refinement 
level is reached in a given cell, no further refinement is performed and a 
sink particle can be formed when the density in this cell exceeds a threshold 
density, $n_{\rm thr}$, among other standard tests. The sink particles can
then accrete mass from their surroundings, increasing their mass 
\citep{Federrath+10}.

Our numerical setup is very similar to that of \cite{Juarez+17}
and \citet[][simulation labeled run 22]{BP+15}.\footnote{Although 
these works are aimed at simulating compact and massive cores and 
the latter authors use a different numerical scheme (Lagrangian 
smoothed particle hydrodynamics) and do not consider magnetic fields.} 
We consider a numerical box of $L=13.5$~pc per side filled with molecular 
gas with number density of $n_0 = 100 \, \mathrm{cm}^{-3}$. Therefore, 
the box contains $1.4\times 10^4 ~\Msun$ (taking a mean molecular 
weight of $\mu=2.3$) of molecular gas homogeneously distributed. 
The initial free-fall time for this configuration is 
$t_{\rm ff} = 3.4$~Myr. The simulation is 
isothermal, with temperature of 10 K.
The isothermal sound speed is thus $c_s =0.19 \, \kms$.

The initial velocity field corresponds to a supersonic decaying
turbulence, so no forcing at later times is imposed. Following the
prescription by \citet{Stone+98}, we included pure rotational velocity
power spectrum with random phases and amplitudes that peak at
wavenumbers of $k= 4\pi/L$. The resulting initial velocity field is
thus a solenoidal supersonic turbulent fluid, with an rms Mach number
of $\mathcal{M}_{\rm rms}=8$.

To properly follow the gravitational fragmentation and collapse of
dense gas, we fulfill the Jeans criterion by resolving the local Jeans
length for 8 grid cells. We allow the code to refine up to reach a maximum 
resolution of $\Delta x_{\rm min} \sim 1.3 \times 10^{-2} \,
\rm{pc}$, and we set the threshold density for sink formation at 
$n_{\rm thr} \sim 10^5 \, \ppcc $. We evolve the simulation for roughly one 
free-fall time.

The magnetic field, of strength $3.7 \, \mG$, is initially uniform 
along the $x$-direction, since the initial orientation does not 
significantly change the results in gravo--turbulent MHD simulations 
\citep[see, e.g.,][]{hennebelle2013}. This initial 
magnetic field strength is within the observational rage in diffuse 
clouds \citep[$\sim$0.1-10 $\mu$G with a mean value of 
$\sim$5-6 $\mu$G;][]{Crutcher+10}. Note that our box is, thus, magnetically 
supercritical, with a mass-to-flux ratio of $\sim$7.0 times the critical value, 
$\mu_\mathrm{crit}=(4 \pi^2 G)^{-1/2}$ \cite[see e.g.,][]{NN78}, 
and prone to hierarchical gravitational collapse as soon as the 
initial turbulence is dissipated.
It should be noticed that although this intensity of magnetic 
field seems to be low for a protostellar core, it is only the initial 
value. As collapse proceeds, gravity is dragging the magnetic field lines, 
increasing the intensity of the field. As a result, our magnetic field 
intensities in the subregion we are analyzing after almost one free-fall 
time have a mean value of $\sim 50\mu$G, with a standard deviation of 
$\sim 35\mu$G. These values are in agreement with the values reported by  
\citet[][]{Crutcher+10}.

\begin{figure*}
	\includegraphics[width=1\hsize]{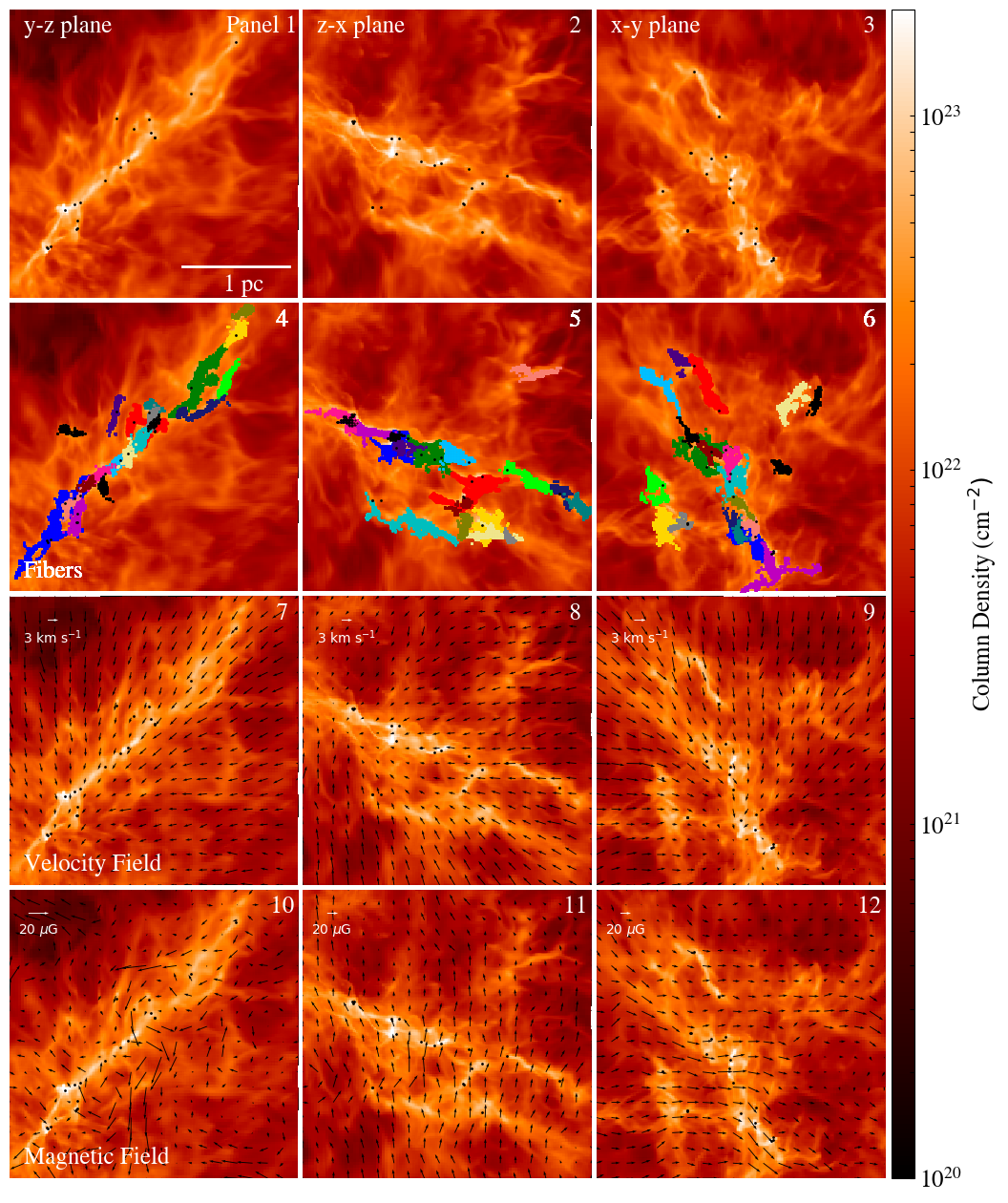}
	\caption{Column density maps of the studied filament integrated along the 
	$x$, $y$, and $z$--direction (left, middle, and right column, respectively). 
	In the second row, the detected fibres in each direction are showed. 
	In the third and fourth row the velocity and the magnetic field vectors are 
	drawn (the vector scale is shown in the upper--left corner in each panel).} 
	\label{fig:fil}
\end{figure*}

\section{Filament and velocity coherent structure finding} \label{subsec:ppv}

To understand the origin of fibres, we analyse either the PPV cubes, and apply 
to them a FoFv algorithm \citep[see, e.g.,][]{fof, hacar2013}, and the PPP cubes, 
applying to them the {\tt DisPerSE} method \citep{disperse1, disperse2}.

\subsection{PPV maps}

Although ideally one would post-process the numerical simulations 
with radiative transfer calculations of the line emission, as in
\cite{smith2016}, we note that radiative transfer simulations have
their own uncertainties, such as: how much shielding of CO from
UV-dissociating radiation is due to local or global gas and dust; 
whether the heating by protostars is significant; or whether the 
commonly used Large Velocity Gradient method is applicable to 
low-velocity or subsonic structures. Therefore, not just as an 
initial step, but also with the aim to understand what an observer 
using a really optically thin line is detecting, we simply assume 
that velocity coherent dense structures will also be seen as 
velocity coherent structures in line emission, although with 
some scaling in intensity. We therefore construct PPV cubes as 
density-weighted velocity histograms along each direction, 
equivalent to an optically thin line with a constant source function.

As a first step to construct the PPV data, we map the density and 
velocities in a regular grid at the maximum resolution of the simulation. 
Then we make a synthetic PPV cubes as density--weighted velocity
histograms at every position $x_1, x_2$, in which we store a 
synthetic intensity, defined as
\beq
I(x_1,x_2,v_3) = \sum_{x_3} n(x_1,x_2,x_3:v_3) 
\label{eq:intensity}
\eeq
where $x_1$ and $x_2$ are any pair of cartesian coordinates, and $x_3$
the corresponding 3rd coordinate, perpendicular to the previous ones,
over which we will integrate in bins of velocity $v_3$ in that
direction. We consider only cells with number density above $10^3 \, \ppcc$,
from which we expect most of the   molecular emission, mimicking 
the emission of an intermediate density tracer like C$^{18}$O. 
We have chosen a velocity channel width of 0.07$\kms$, as
\cite{hacar2013}, in order to allow detailed comparison.

\subsection{Friends--of--Friends in velocity}

Following \citep[][hereinafter H13]{hacar2013}, we used 
a Friends--of--Friends in velocity (FoFv) algorithm to find coherent 
structure in the PPV data cubes. This is a similar algorithm to 
Friends--of--Friends \citep{fof}, but the 3rd coordinate is the velocity
in the line of sight. We chose to follow H13 as close as possible,
and so, we use the same parameters to build the PPV diagrams and find
structure within them. As H13, we followed a 3 step algorithm:
we first identify the high intensity points choosing those points having
6 times the root mean square value of the intensity, $I_{\rm rms}$, in the 
PPV cube, then using the first selection, we search for friends of friends 
using twice the spatial resolution in distance, and finally, points with
intensities lower than 6$\times I_{\rm rms}$, but still larger than 
$3 \times I_{\rm rms}$, are considered (see section 6 and Fig. 10 in H13).

This method allows us to identify velocity--coherent structures, which we
will refer as ``fibres", in the filaments of the simulations.
Note that, in contrast to observations, we are able to trace fibres in
the 3 projections, and store the actual location of every cell in the
PPP space.

For consistency, we use the same resolution in PPV space that H13, 
along with the same FoFv parameters. The only free parameter is 
the density threshold to find the densest PPV cells, which we chose 
to 6$\times I_{\rm rms}$, although we get roughly the same number of
fibres in the range of $\sim$5-8$\times I_{\rm rms}$. On the other hand, 
increasing or decreasing the minimum intensity for which we find friends 
results in thinner or wider fibres in the velocity coordinate.

\subsection{3D Filament finding}

We use {\tt DisPerSE} 
\citep[Discrete Persistent Structures Extractor;][]{disperse1, disperse2} 
in order to find filamentary structures from the 3D density field based in 
topological considerations.

{\tt DisPerSE} is based on the discrete Morse Theory. This algorithm find 
filaments by locating critical points (where the gradient of the density 
field is zero) and the skeleton of the filament is building by connecting 
maxima and saddle points along integral lines (curves tangent to the gradient 
filed). The important free parameter is the so-called persistence threshold 
(the absolute difference value between two critical points, e.g. a maximum 
and a saddle), which we set to $10^4 \, \ppcc$. This value is roughly an 
order of magnitude greater than the mean density of the sub--box containing 
the studied filament (white squares in Fig. \ref{fig:cloud}), which prevent 
us of picking up less significant structures from the density noise.

\section{Results} \label{sec:results}

\subsection{General evolution}

The global evolution of our box is described as follows.\footnote{A movie 
of the evolution of the simulation can be seen in the online material.} 
As commented previously, we started with a 13.5 pc size box with constant 
density field at 100~cm\alamenos 3, and constant magnetic field parallel 
to the $x$ direction. The initial velocity field fluctuations induce shocks 
that rapidly dissipate. Since we do not have turbulence forcing, 
the initial turbulence (with Mach number of 8) dominates only 1/5$^{th}$ 
of the evolution time.
The total evolution of the box is around one free-fall time, equivalent to 
$\sim$~3.4~Myr.\footnote{Although the free-fall time will be a better 
reference in order to make scalable the simulation to any regime, in what 
follows, we will use Myr for a more physical reference, but it should be 
noticed that since the simulation is isothermal, the time is scalable 
according to the adopted mean density} As soon as most of the turbulent 
energy is dissipated ($t\sim 0.75$~Myr) the turbulent induced filaments 
start accreting material, producing well--defined filaments which in turn, 
merge each other to form larger structures by $t\sim 1.78$Myr. 
By $t=2.6$~Myr ($\sim 0.8$ free-fall times), there are clearly several 
high column density filaments, and the whole material falls towards the 
center of the box. At this time, some traces of the boundary conditions 
(a box) are clearly seen: the collapsed gas  exhibits a squared box, 
with material piling up somehow nearly along the diagonals of the box, 
due to the gravitational focusing described by \citet{HartmannBurkert07}.

In Fig. \ref{fig:cloud} we show the column density of the central region 
(of 7.25 pc per side) of our computational box for the three projections 
of our numerical simulation at $t \sim 3.1$ Myr (or $\sim0.9 \, \tff$), 
0.5 Myr after the onset of star formation. Although several filaments 
are seen in the box, we focused our efforts in a single region (enclosed 
in white squares in Fig. \ref{fig:cloud}), which exhibits an extended 
filament in the $x$--projection, but when seen in the other projections, 
the structure appears to be more complex, with several sub--filaments. 
This morphology allows us to understand the role of ``fibres" compared 
with the physical structure of the subfilaments.

Fig. \ref{fig:fil} shows a zoom--in of the region
containing the filament. The left, middle and right panels show the 
$x$, $y$, and $z$ projections (in the $y-z$, $z-x$ and $x-y$ planes, 
respectively). The upper row is just the column density field, along 
with the sink particles as black points.  The second row shows the 
fibres found in each projection, as discussed below (\S \ref{sec:real?}), 
one color per fibre. 
The third and four rows show the velocity (panels 7-9) and magnetic 
(panels 10-12) fields projected in the plane of the corresponding image. 
In both cases, the projected arrows represent the mean value 
(volume--weighted) of the vector field along each line of sight. The 
velocity scale is shown in the upper left corner.

There is a couple of features to note in this figure. As the evolution 
of the simulation shows, the whole box is collapsing towards the center 
(see Fig.~\ref{fig:cloud}). Thus, although the gas is falling in, it locally 
does it through the filament, while the filament itself is moving also to 
the center of the box. In that sense, the filament is not formed by 
large-scale turbulent streams, but by gravitationally focused flows. 
Looking at the 3 projections (panels 7, 8 and 9 of Fig. \ref{fig:fil}), it can be seen that 
the flows towards the filament, is somehow oblique to it. In the frame 
of reference moving with the filament, however, it is clear that the gas 
is falling almost radially into the filament (see \S\ref{sec:skeleton}). 
On the other hand, although the magnetic field is also more or 
less perpendicular to the filament in the $z-x$ and the $x-y$ planes, in 
the $y-z$ plane looks substantially more chaotic. This is because the 
original orientation of the field was along the $x$ axis, and as the
simulation evolves and the material falls down into the center of the 
computational domain, it drags the field lines perpendicular to the $x$ axis.

\begin{figure*}
	\minipage{0.5\textwidth}
	\includegraphics[width=\linewidth]{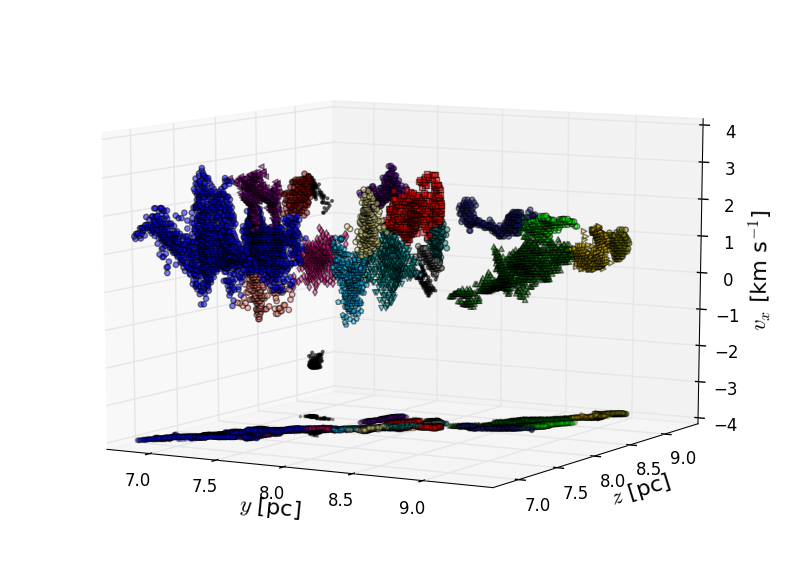}
	\endminipage\hfill
	\minipage{0.5\textwidth}
	\includegraphics[width=\linewidth]{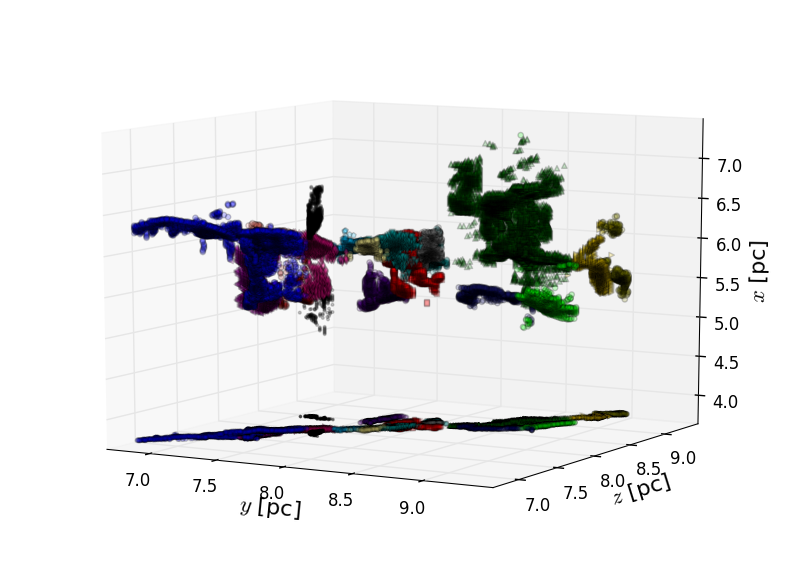}
	\endminipage\hfill
	\caption{ {\it Left:} fibres found along the $x$--LOS showed in the 
	PPV space. The colors represent independent fibres. {\it Right:} 
	fibres mapped in the corresponding PPP diagram. The color scale is 
	the same that for the projected fibres shown in Figure \ref{fig:fil} 
	(panel 4).}
    \label{fig:ppv-x}
\end{figure*}

\subsection{Are fibres real objects?} \label{sec:real?}

We begin by describing results from a projection along the $x$ 
line--of--sight (LOS; $y-z$ plane; first column in Figure \ref{fig:fil}).  
In this projection, the column density exhibits the narrowest filamentary 
structure. Using our FoFv algorithm, we identify 20 separate 
velocity-coherent fibres in the left panel of Fig. \ref{fig:ppv-x}. 
These fibres are also projected in the column density map of 
Fig. \ref{fig:fil} (Panel 4). The results bear a close qualitative 
resemblance to the velocity--coherent groups identified in Figure 13 
of H13. The velocity range of the observed groups span $\sim 2 \kms$, 
comparable to what we find here. Note that in Panels 4--6 of 
Figure \ref{fig:fil} we show the projection of all elements of the 
fibres, whereas H13 only show the central axis of each fibre in 
their Fig. 12.

The right-hand panel of Figure \ref{fig:ppv-x} identifies the
positions of the (color--coded) coherent velocity groups in real 
(PPP) space. We note that while some of the groups appear to be 
roughly co-spatial, clearly there are others that are spatially 
separated but are superposed along the line of sight.

\begin{figure*}
	\minipage{0.5\textwidth}
	\includegraphics[width=\linewidth]{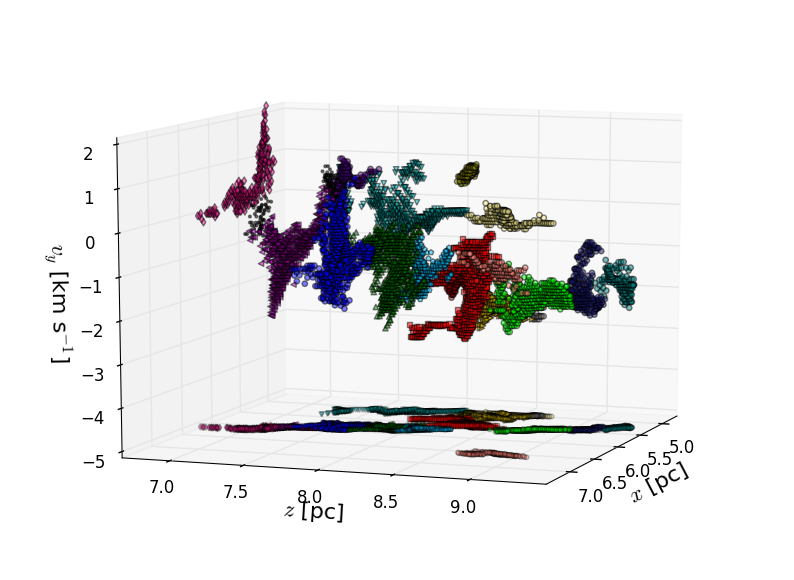}
	\endminipage\hfill
	\minipage{0.5\textwidth}
	\includegraphics[width=\linewidth]{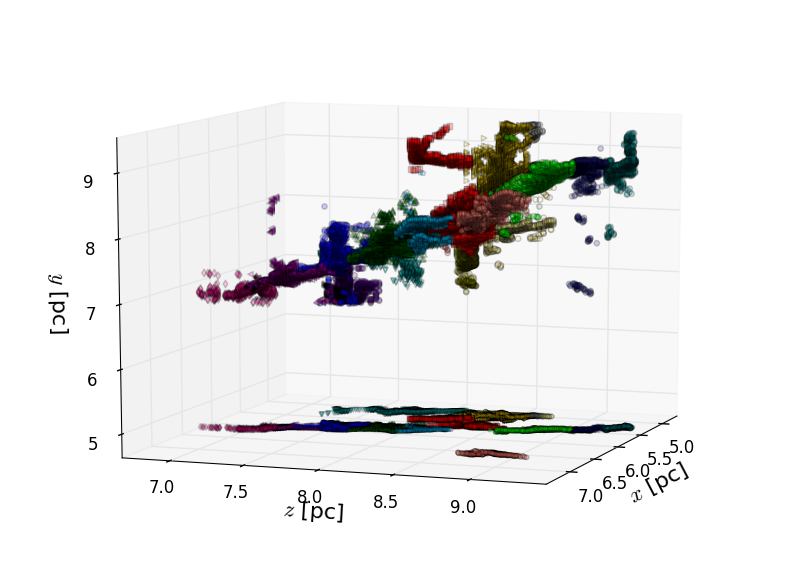}
	\endminipage\hfill
	\caption{ Same that Fig. \ref{fig:ppv-x}, but for fibres found in the $y$--LOS. The color scale is the same that for the projected fibres 
	shown in panel 5 of Figure \ref{fig:fil}.}
    \label{fig:ppv-y}
\end{figure*}

\begin{figure*}
	\minipage{0.5\textwidth}
	\includegraphics[width=\linewidth]{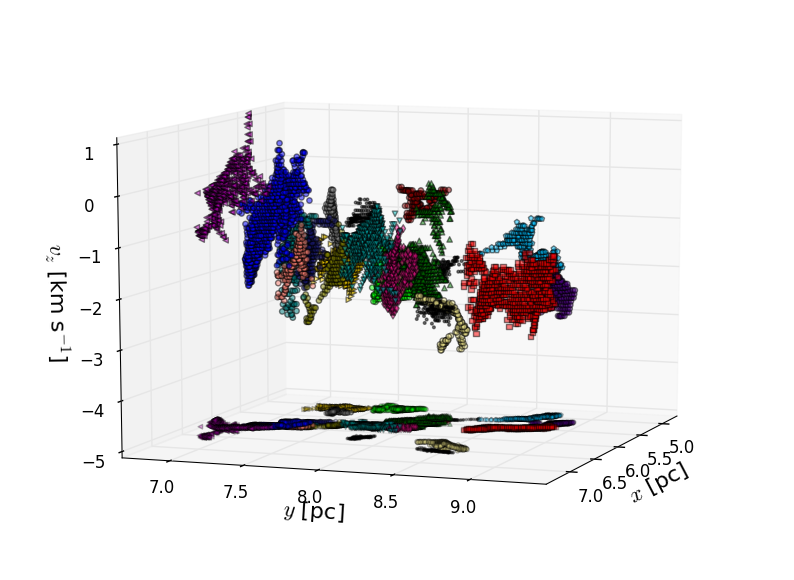}
	\endminipage\hfill
	\minipage{0.5\textwidth}
	\includegraphics[width=\linewidth]{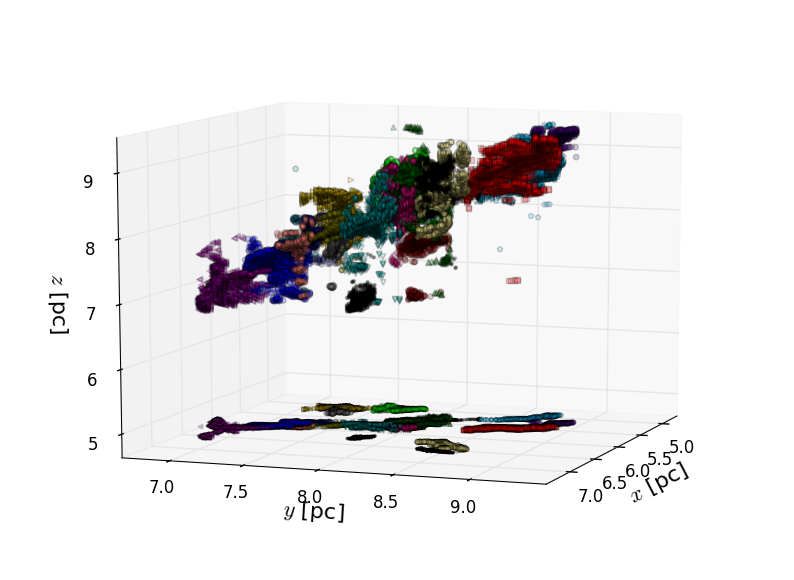}
	\endminipage\hfill
	\caption{ Same that Fig. \ref{fig:ppv-x}, but for fibres found in the 
	$z$--LOS. The color scale is the same that for the projected fibres 
	shown in panel 6 of Figure \ref{fig:fil}}
    \label{fig:ppv-z}
\end{figure*}

\begin{figure*}
	\minipage{0.5\textwidth}
	\includegraphics[width=\linewidth]{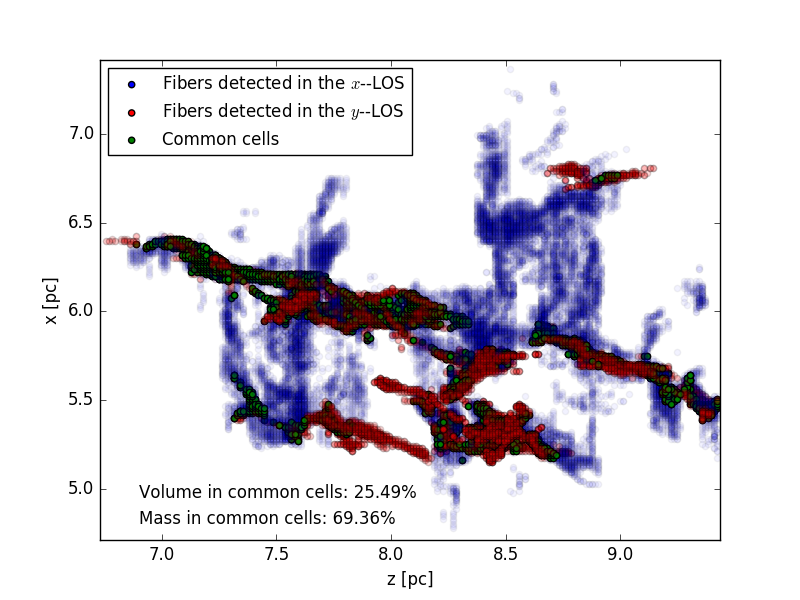}
	\endminipage\hfill
	\minipage{0.5\textwidth}
	\includegraphics[width=\linewidth]{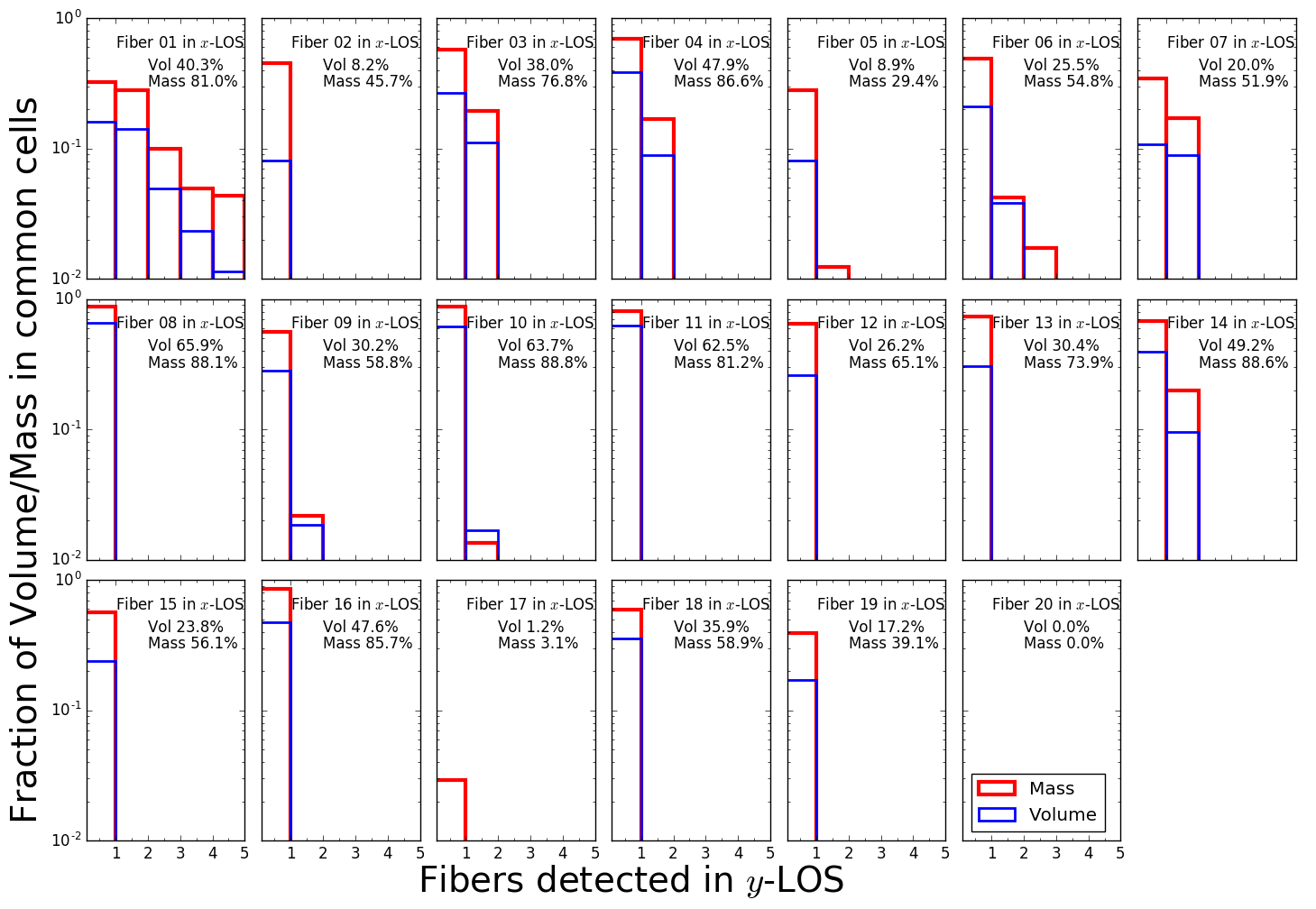}
	\endminipage\hfill
	\caption{ {\it Left:} Projection in the $z-x$ plane of the
          fibres found along the $x$-- and $y$--LOS (blue and
          red symbols, respectively). The common points are colored in
          green. {\it Right:} Fraction of volume and mass (blue and red bars, 
          respectively) that a given fibre in the $x$--LOS shares with fibres 
          detected in the $y$--LOS.}
    \label{fig:fib_intersec}
\end{figure*}

Figures \ref{fig:ppv-y} and \ref{fig:ppv-z} show the corresponding
results for the other two projections (see also their superposition 
on the column density maps in Panels 5 and 6 of Figure \ref{fig:fil}). 
The spatial distinction between the various coherent velocity components 
is clearer in these projections.

A common feature of the PPV groups is the ``sawtooth'' structure, 
with opposing velocity gradients over scales of a few tenths of a pc.
Similar sawtooth patterns are also seen in the observations of H13.
In several cases we can correlate the positions of the velocity
gradients with cores and/or sinks; this suggests that the sawtooth 
pattern traces gravitational acceleration toward mass concentrations 
\citep[see, e.g., ][]{Kuznetsova+17}.

A crucial question is whether the ``fibres" are actual 
coherent three--dimensional objects with subsonic internal velocities. 
To address this issue, we correlate the pixels or grid cells associated 
with a fibre in one projection with those of another. The left panel of 
Figure \ref{fig:fib_intersec} shows the resulting overlaps (green dots) 
of fibres found along $x$-- and $y$--LOS (blue and red dots, respectively). 
We find that only $\sim$25\% of the cell fibres volume are identified in 
both projections, which contains $\sim$69\% of the fibres mass. We find 
similar numbers when comparing fibres in the other projections, which imply 
that only the densest gas is systematically located in fibres regardless 
of the LOS. 

The right panel of Figure \ref{fig:fib_intersec} shows the volume 
and mass fraction (blue and red bars,  respectively) that one fibre in the 
$x$--LOS projection shares with other fibres identified in the $y$--LOS. 
Thus, $n$ bars in a given histogram means that this fibre found in the $x$ 
projection shares material with $n$ fibres in the $y$ projection.

From this figure, we notice that not all the volume (mass) that 
generate a fibre in one LOS necessarily participate in the formation of a 
fibre in another LOS. In fact, most of the fibres share a small fraction of 
volume with only one or two other fibres found in the other LOS. Furthermore, 
it is interesting to note that the mass fractions are a bit larger than the 
volume fractions, since the overlapping regions are the densest parts of the 
filament. It should be noticed, never the less, that in none of the cases all 
the volume (mass) of one fibre goes completely into a single other fiber 
in another projection.

Finally, we notice also that there are few fibres with few common pixels or 
no counterpart in different LOS (e.g., fibres 17 and 20 in the right panel 
of Fig. \ref{fig:fib_intersec}). These results 
suggest that great caution should be applied assessing the reality of 
``coherent--velocity'' structures, i.e., fibers seem not to be 
single 3D objects.

\subsection{The skeleton as a real 3D structure; kinematics of the filaments}\label{sec:skeleton}

Another approach to understand the nature of the filaments is to analyse their 
actual 3D data. Although this is not possible in observations, it 
turns out to be quite instructive in the simulations. 

\begin{figure*}
	\centering
	\begin{minipage}{1.0\textwidth}
		\centering
		\includegraphics[width=0.49\textwidth]{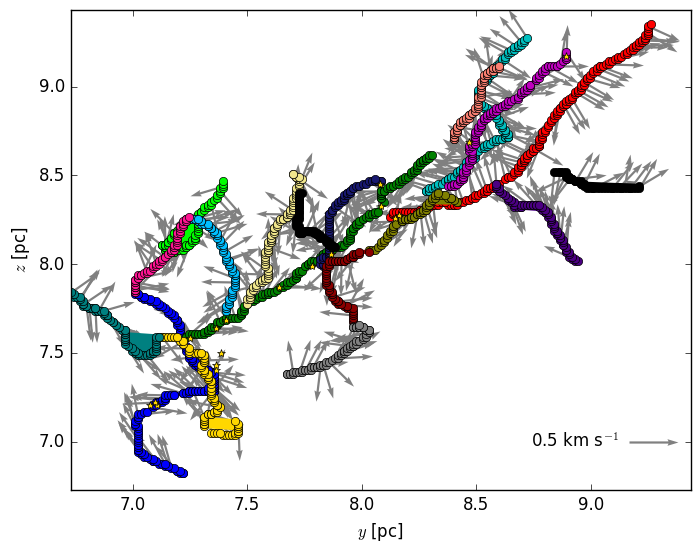}
		\includegraphics[width=0.49\textwidth]{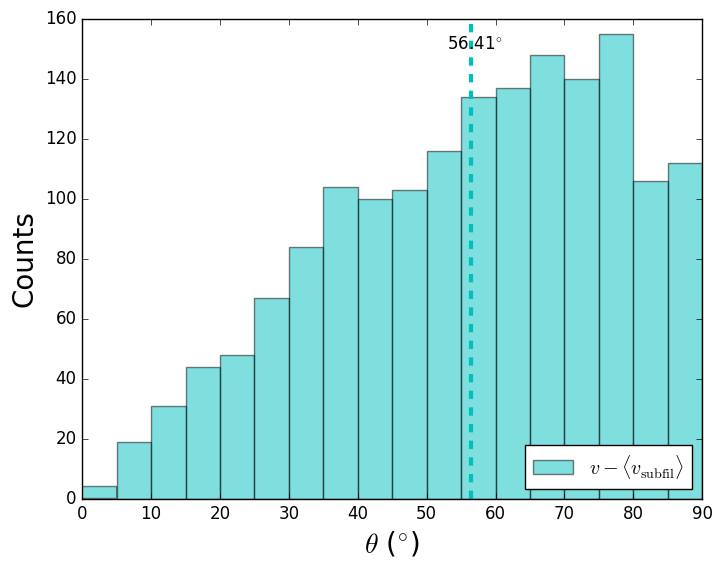}
		\caption{{\it Left:} 3D subfilaments found with {\tt DisPerSE}
	projected in the $y − z$ plane. The arrows are the projected 
	velocity field in each point of the subfilaments in 
	the frame of reference of each subfilament. The vector scale 
	is shown in the lower left corner. The yellow stars correspond 
	to the projected positions of the sink particles. {\it Right:} Histograms 
	of the angle between the skeleton and the 3D velocity field. The vertical 
	dashed line represent the average.} 
		\label{fig:skeleton}
	\end{minipage}\hfill
\end{figure*}

The left panel of Figure~\ref{fig:skeleton} shows the 
segments or ``subfilaments" found by {\tt DisPerSE} in our data, and the 
velocity field vectors, computed now in the frame of reference of the 
filament, at the position of the skeleton. In the right panel of 
Figure~\ref{fig:skeleton} we furthermore compute the histogram of the 
angles between the velocity field in the frame of reference of the filament,
and the local direction of the skeleton.

Two features arise naturally: first of all, the skeleton (in projection) 
occupy nearly the same loci of the ``fibres" (see panel 4 of 
Figure~\ref{fig:ppv-x}).  Second, as commented in the previous section, 
in the frame of reference of the filament, the velocity vectors show a 
tendency to be perpendicular to the filament.

\begin{figure*}
	\centering
	\begin{minipage}{1.0\textwidth}
		\centering
        \includegraphics[width=0.49\textwidth]{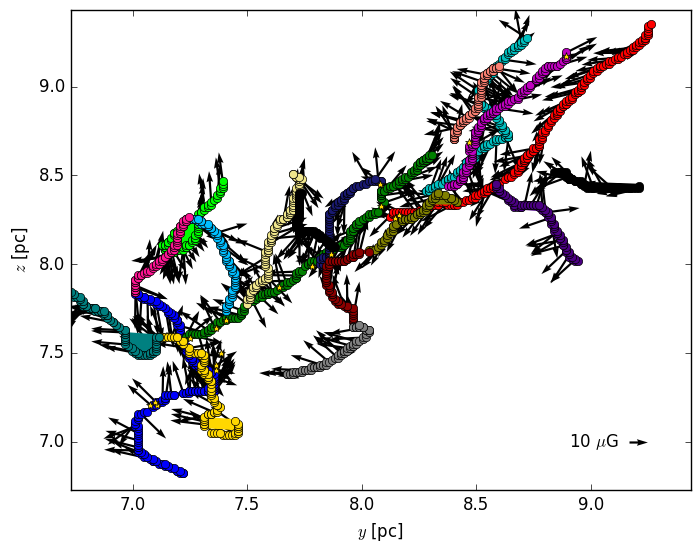}
        \includegraphics[width=0.49\textwidth]{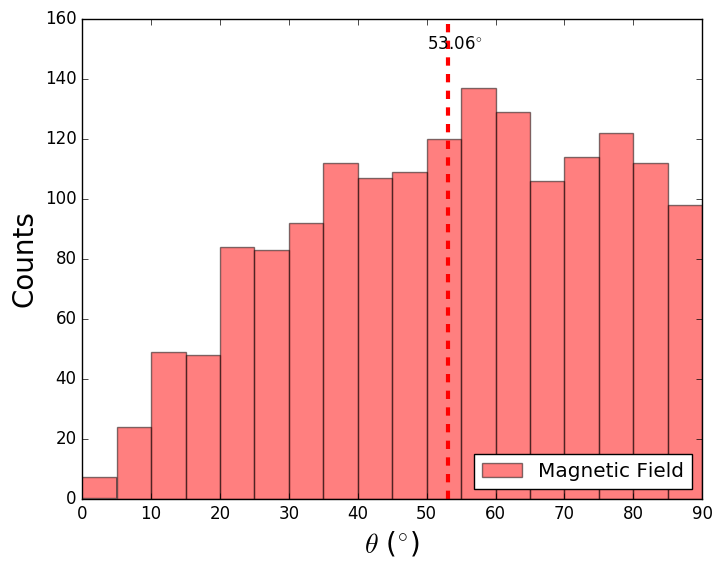}
		\caption{{\it Left:} 3D subfilaments found with {\tt DisPerSE}
	projected in the $y − z$ plane. The arrows are the projected 
	magnetic field in each point of the subfilaments. The vector scale is shown 
	in the lower left corner in each panel. The yellow stars correspond 
	to the projected positions of the sink particles. {\it Right:} Histograms 
	of the angle between the skeleton and the 3D magnetic field. The vertical dashed 
	line represent the average.}
		\label{fig:skeleton_mag}
	\end{minipage}\hfill
\end{figure*}

Looking at this figure one may conclude, at first glance, that each 
subfilament is moving almost perpendicularly to its main axis, and thus, 
that they should be the result of large-scale turbulent shock fronts. 
However, as shown in Figure~\ref{fig:fil} 
(see also the online material),
this is not the case: the filament is falling as a single entity towards 
the center of the box. Thus, what the velocity arrows in 
Figures~\ref{fig:fil} and \ref{fig:skeleton} are showing us is that, 
in the frame of reference of the filament, the gas is falling into it, 
almost perpendicularly, while the whole filament falls into the 
gravitational potential well. 

The filament orientation is the result of the boundary 
conditions: it reflects the overall non-spherical initial geometry of the 
cloud \citep[e.g.,][]{Kuznetsova+15}, such that the density field suffers from 
gravitational focusing at the corners of the box. As a result of the interplay 
between the initial turbulence and the geometry, filaments along the diagonals 
of the box are formed rapidly. In our case, this effect allows the formation of 
a filament along the large dimension (the diagonal of our cube).  At the same 
time, as it is formed, the filament falls down into the global gravitational 
potential well at the center of the box.
In this process, the filament is constructed or fed from the large scales, mostly 
perpendicularly, due to self-gravity, while it falls down into the global potential 
well.

We have also computed the orientations between the magnetic field and the skeleton. 
In Fig.~\ref{fig:skeleton_mag} we show the skeleton but now with the magnetic field 
vectors, and the corresponding vector--skeleton histogram. In this case, the magnetic 
field looks slightly less aligned preferentially with the subfilaments, with a wider 
histogram and a less pronounced peak than the velocity histogram.

\begin{figure}
	\centering
	\begin{minipage}{.49\textwidth}
		\centering
		\includegraphics[width=\linewidth]{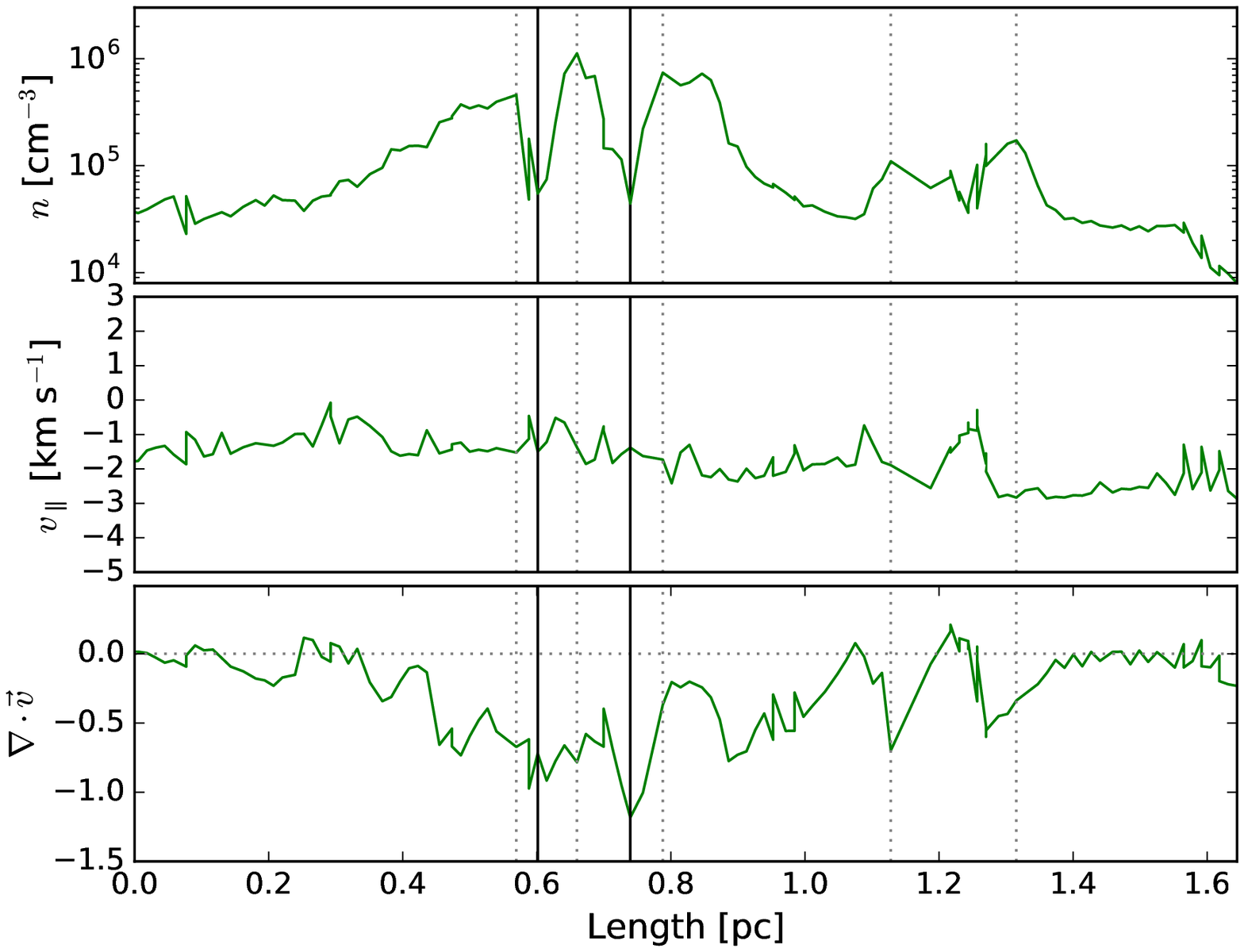}
	\end{minipage}\hfill
    	\begin{minipage}{.49\textwidth}
		\centering
		\includegraphics[width=\linewidth]{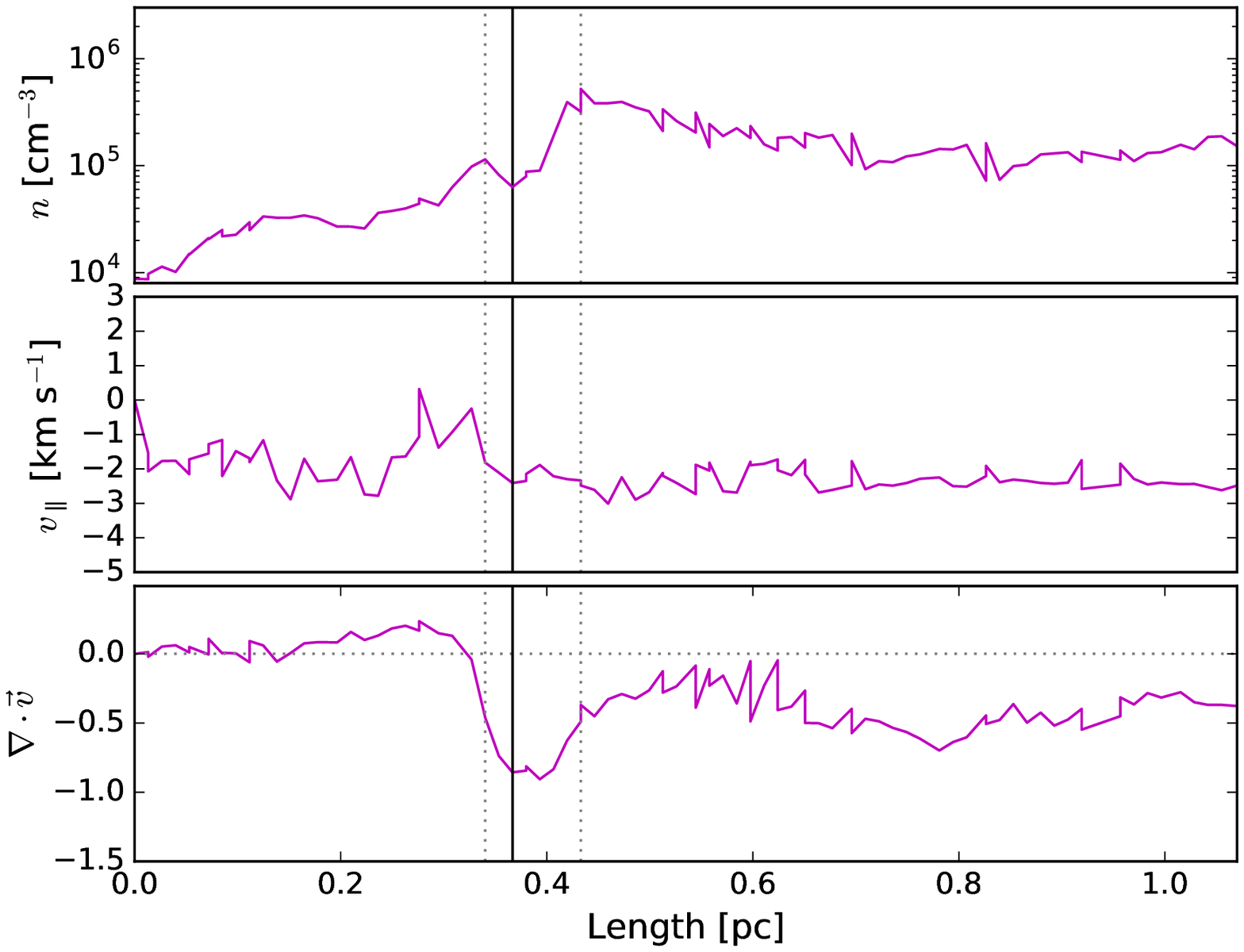}
	\end{minipage}\hfill
	\begin{minipage}{.49\textwidth}
		\centering
		\includegraphics[width=\linewidth]{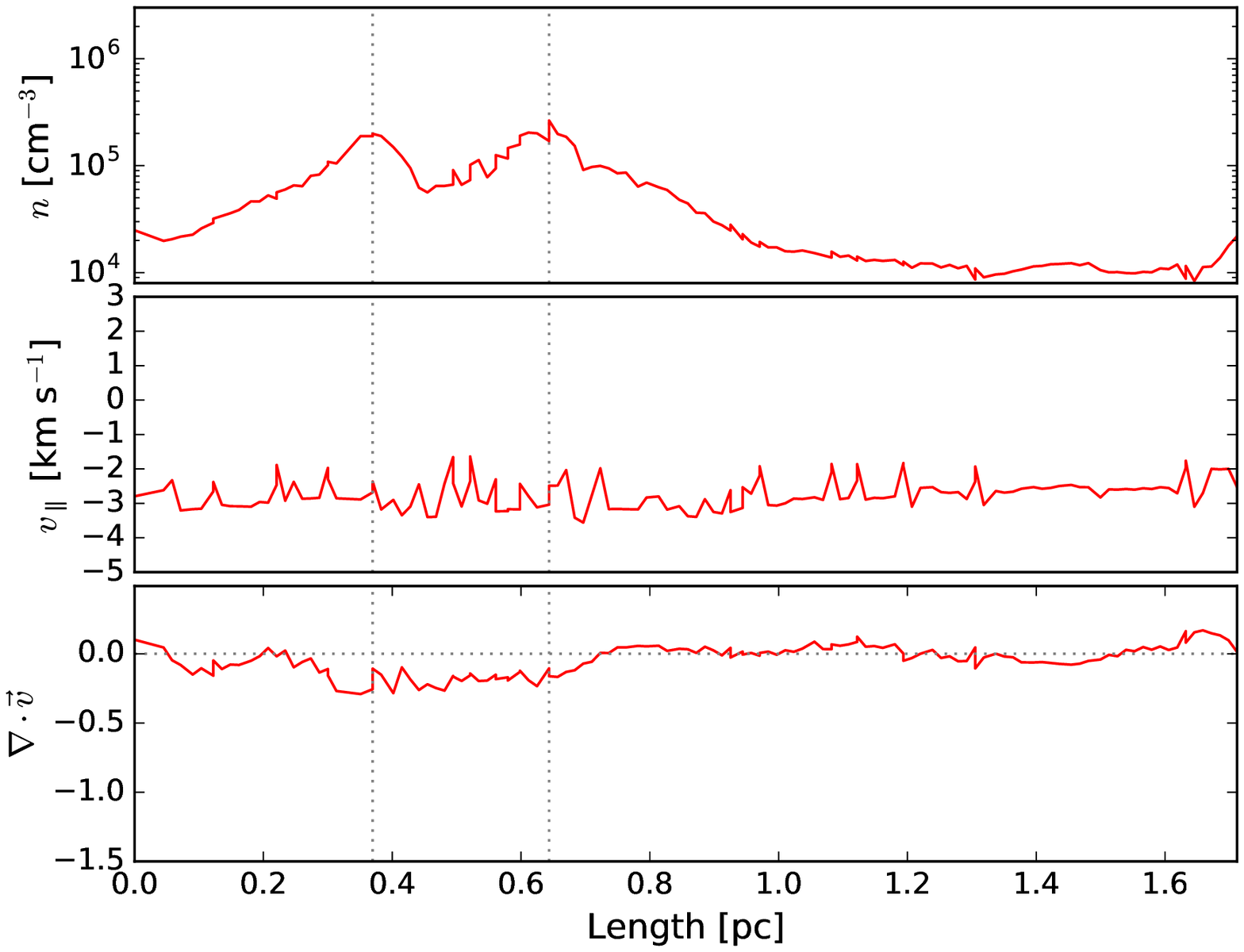}
	\end{minipage}
		\caption{Each sub--figure shows the density, the parallel
                  velocity component, and the 3D divergence (top, middle, and
                  bottom panels) of the green, magenta, and red
                  sub--filaments shown in Fig. \ref{fig:skeleton}.
                  The vertical lines represent the positions of sink particles 
                  (solid lines) and some local density maximums (dotted lines).}
		\label{fig:profiles}
\end{figure}

Analysing the velocity field along the skeleton, in the local frame of 
reference, is instructive too. To do so, we have plotted, in 
Figure~\ref{fig:profiles}, from top to bottom panels, the density profile, 
the velocity field profile component parallel to the skeleton, and the 
divergence, $\nabla\cdot\vec{v}$, of the velocity field along the 
skeletons of 3 subfilaments found with {\tt DisPerSE}. The colors of the 
plots also denote the subfilament in Figure~\ref{fig:skeleton} (left panel): 
the green, the magenta, and the red subfilaments. The first interesting 
feature to notice is that the divergence of the velocity field along
the skeletons is predominantly negative, implying that the skeletons 
exhibit converging motions towards them as a fundamental characteristic. 
The negative divergence indicates that the subfilaments are accreting mass 
from their surroundings. 

Another interesting feature of Fig.~\ref{fig:profiles} is that at the 
positions of the sinks (denoted by vertical solid black lines), the 
divergence of the velocity field is more strongly negative due to the 
sink gravity, and the filament itself has a local minimum of the density 
because the mass has been accreted by sink particles. Finally, the third 
interesting feature is that the local density maximums do not correlate 
with sharp shocks in $v_{||}$, implying that, along the filaments, local 
excesses of density are produced mainly by accretion perpendicular to the 
filament, not by shocks inside and along the filament itself.

\section{Discussion} \label{sec:discussion}

We have shown that simulations of clouds with turbulent--decaying
substructure and with evolution driven by gravity, naturally produces
both filaments and subfilaments which, when seen along certain lines
of sight, resemble the velocity--coherent fibres of H13 and
\cite{tafalla2015}. Most of these sub-filaments are distinct
objects in space where gas has been compressed and/or shocked.  The
subfilaments arise independently of the main filament, unlike the
fray--and--fragment scenario and may exhibit a spoke--like structure as
they accelerated toward the filament, especially if there are mass
concentrations present (i.e., cores and sinks).

\subsection{Comparision with previous works}

The previous work closest to ours is that of \cite{smith2016}, who
considered turbulent cloud evolution without magnetic fields.  We find
that the overall morphology is not strongly affected by the inclusion
of a modest magnetic field, although the main filament that results is
somewhat more linear.

\cite{smith2016} analysed their
results in a qualitatively different manner than in this paper. They
applied the {\tt DisPerSE} algorithm
to identify (sub)filaments in volume
density space, and then considered the velocity fields of these
structures, decomposing the three-dimensional velocity field into
flows along and perpendicular to the filament.  While they show that
there are multiple velocity components along certain lines of sight
through the filament region, they do not make the identification of
their sub-filaments in PPV space and so the results are not directly
comparable to those of H13 (e.g., their Figure 11).  We also differ
somewhat in our interpretation of the simulations; while Smith et al.
talk about collecting sub-filaments due to ``large-scale,
low-wavenumber modes and gravitational collapse'', we would interpret
the motions as predominantly driven by gravitational acceleration 
an indication of which is the spoke structure seen around ``hubs'' of
mass concentration (cores and sinks).

\cite{clarke2017}, on the other hand, investigated the effects of 
turbulent velocity fields on a filament.  While they found that 
supersonic turbulence could create elongated structures within the 
filament, suggestive of the fray and fragment scenario for producing 
fibres, the initial setup - a cylindrical computational volume of 
radius 1 pc and length 3 pc, with a centrally peaked initial density 
distribution and radial inflow - precludes any investigation of fibre 
production on larger scales independently of or concurrently with 
filament formation, as we do here.  Moreover, the negative divergences 
shown in Figure \ref{fig:profiles} indicates that the inflow is not 
smooth, but structured, which can also affect the development of 
fragmentation.

We emphasize that some of the ``coherent'' velocity structures found 
along one projection are not found to be such in other projections. 
Thus, few fibres may simply be projection effects enhanced by the 
complex velocity field.  H13 made a distinction between ``fertile'' 
(i.e., core-forming) and ``sterile'' fibres. Our simulations and 
analysis suggest that some of the sterile fibres might simply be 
projection effects of structures that are not really coherent in 
three dimensions. The ``fertile'' fibres may simply be those which 
are the densest or have the most mass - which, as we have shown are 
more likely to be coherent structures in space.

\section{Conclusions} \label{sec:conclusion}

We have analysed a numerical simulation of a self-gravitating molecular 
cloud including sink formation and magnetic fields to attempt to explain 
the nature and origin of apparent velocity--coherent structures - called 
fibres - in observations of star-forming filaments by \cite{hacar2013}. 
We use similar fibre--finding algorithms to identify these structures in 
three different PPV projections.  While we find that fibres identified in 
this way are ubiquitous, we also show that some fraction of these objects 
are not coherent {\it physical} structures, as shown by our finding that 
what are identified as fibres in one projection are not the same as in 
the other projections.  Only the densest structures maintain their 
coherence in space. Overlapping of distinct physical structures seen in 
projection accounts for the appearance of multiple velocity components 
in the same region of the main filament.

In our simulation, there is global gravitational collapse because the 
cloud, especially once the input turbulence decays, is subvirial. The 
effect of the initial turbulence is mostly to produce density concentrations 
which result in a complex velocity field due to gravitational acceleration. 
Filaments (and subfilaments) form by accretion/infall from the surrounding 
medium, driven by gravity, not by turbulence. Seeded by the turbulent 
fluctuations, filaments, cores, and sinks form, producing structures -- 
in some cases physically coherent, in others simply the result of projection 
of independent regions -- which, when observed in PPV space, can be 
interpreted as ``fibres'' in the sense of H13.

Analysing 3D structures obtained with {\tt DisPerSE}, we also find that 
the skeleton is mainly accreting material radially from its surroundings, 
hierarchically collapsing and forming stars, and falling globally into 
the gravitational potential well of the simulation, at the same time. 
The accretion of material is along the magnetic field lines, which are 
oriented preferentially perpendicular to the skeleton (with a mean angle 
of 53 degrees), and are being bent and dragged by the velocity field. 

Finally, our work has implications for the ``fray and fragment" scenario.
Velocity--coherent structures naturally appear in our hierarchically and 
chaotically collapsing simulation, in which the ``seeds" giving rise to
the fibres were formed some distance away from the main filament by turbulence,
and then concentrate spatially mainly by gravity to form the coherent filament 
\citep[see also][]{smith2016,moeckel2015}. This is the opposite process
envisaged in the fray and fragment scenario; thus the presence of 
velocity--coherent structures does not necessarily demonstrate that the 
``fray and fragment" scenario is operating.

\section{Acknowledgments}

We thank the anonymous referee for a helpful and 
constructive report that improved the clarity of this paper.
We thankfully acknowledge Mario Tafalla for sti\-mu\-la\-ting discussions.
J.B.P. acknowledges UNAM-PAPIIT grant number IN110816, and to UNAM's 
DGAPA-PASPA Sabbatical program. He also is indebted to the Alexander 
von Humboldt Stiftung for its valuable support. The research of LH was 
supported in part by NASA grant NNX16AB46G. MZA acknowledges CONACyT 
for a postdoctoral fellowship at University of Michigan. We acknowledge 
Christopher Davies, Gilberto Zavala P{\'e}rez, Alfonso H. 
Ginori Gonz{\'a}lez, and Miguel Espejel Cruz for their important 
computational support. The numerical simulation presented here was 
performed at {\it Mouruka} cluster at {\it Instituto de Radioastronom{\'\i}a 
y Astrof{\'\i}sica} (Universidad Nacional Aut{\'o}noma M{\'e}xico), 
acquired through the CONACYT grant INFR-2015-01-252629 to JBP. The 
visualization was carried out with the {\tt yt} software \citep{yt}. 
The FLASH code used in this work was in part developed by the DOE 
NNSA-ASC OASCR Flash center at the University of Chicago. This research 
has made use of NASA's Astrophysics Data System Abstract Service.

\bibliographystyle{mnras}
\input{main.bbl}


\bsp	
\label{lastpage}
\end{document}